# LIGHT-INDUCED DYNAMICS OF LIQUID-CRYSTALLINE DROPLETS ON THE SURFACE OF IRON-DOPED LITHIUM NIOBATE CRYSTALS

Luka Cmok,[1] Nerea Sebastián,[1] Alenka Mertelj,[1]

Yongfa Kong,[2] Xinzheng Zhang,[2] Irena Drevenšek-Olenik[1,3,*]

[1] J. Stefan Institute, SI-1000 Ljubljana, Slovenia
[2] The MOE Key Laboratory of Weak-Light Nonlinear Photonics and International Sino-Slovenian Joint Research Center on Liquid Crystal Photonics, TEDA Institute of Applied Physics and School of Physics, Nankai University, Tianjin 300457, China
[3] University of Ljubljana, Faculty of mathematics and Physics, Jadranska 19, SI-1000, Ljubljana, Slovenia
*irena.drevensek@ijs.si

**Abstract:** We investigated the effect of a photovoltaic field generated on the surface of iron-doped lithium niobate crystals on droplets of a ferroelectric nematic liquid crystalline and a standard nematic liquid crystalline material deposited on this surface. When such assembly is illuminated with a laser beam, a wide range of dynamic phenomena are initiated. Droplets located outside the laser spot are dragged in the direction of the illuminated area, while droplets located inside the illuminated region tend to bridge each other and rearrange into tendril-like structures. In the ferroelectric nematic phase ($N_F$) these processes take place via the formation of conical spikes evolving into jet streams, similar to the behavior of droplets of conventional dielectric liquids exposed to overcritical electric fields. However, in contrast to conventional liquids, the jet streams of the $N_F$ phase exhibit profound branching. In the nematic phase (N) of both the ferroelectric nematic and the standard nematic material, dynamic processes occur via smooth-edged continuous features typical for conventional liquids subjected to under-critical fields. The difference in dynamic behavior is attributed to the large increase of dielectric permittivity in the ferroelectric nematic phase with respect to the dielectric permittivity of the nematic phase.

1. Introduction

Iron-doped lithium niobate crystals ($LiNbO_3$:Fe (LN:Fe), point group 3m) are renowned for their strong bulk photovoltaic effect [1-3]. By illumination with light in the green spectral region, a spatial redistribution of charge carriers occurs in the illuminated area. This redistribution takes place predominantly along the polar axis (c-axis) of the crystal. Consequently, a static space-charge electric field in the range of 10-100 kV/cm is readily generated inside the crystal [4,5]. This field extends into the surrounding space around the crystal in the form of an evanescent field [6]. Via electrophoretic and/or dielectrophoretic forces, the evanescent photovoltaic field can cause trapping and manipulation of small objects deposited on the crystal surface and induce their (re)distribution following the illumination pattern [7,8]. This phenomenon is known as photovoltaic tweezers (PVT). The main advantage of PVTs with respect to the conventional optical tweezers is that even optical beams with relatively low power can cause strong trapping effects. This is beneficial for the manipulation of materials that can be damaged or uninvitedly modified by optical irradiation.

While optoelectronic manipulation of nano- and micro-particles of condensed materials with the PVTs was extensively investigated, reports on experiments with liquid materials are relatively rare. They are mainly focused on the control of droplets of pure water or aqueous dispersions of different compounds based on dielectrophoretic forces associated with the photovoltaic or/and pyroelectric effects [9-13]. Most research achievements in this field are reviewed in [14]. Studies involving unconventional liquids, such as liquid crystals (LCs), are even more scarce. The research with LC materials can be divided into two categories: manipulation of bulk LC layers and manipulation of LC droplets. Research in the former category is aimed mainly at imprinting reconfigurable optical patterns into the LC medium [15-27], while investigations in the latter category deal mainly with rearrangement of the LC droplets in response to photovoltaic or pyroelectric fields [28-31].

All above-mentioned research with LC materials was performed with conventional nematic liquid crystalline compounds. In the traditional nematic phase, rod-shaped molecules are orientationally aligned with respect to each other. Nevertheless, the phase is not polar, as the probabilities that molecular dipoles are pointing in each of the two opposite directions associated with the alignment axis are the same [32]. But, recently, nematic LC compounds exhibiting polar (ferroelectric) orientational order were discovered [33-36]. The ferroelectric nematic LC phase is a proper ferroelectric fluid that is extremely sensitive to external electric fields [37-39]. Therefore, it is naturally anticipated that it should strongly interact with the photovoltaic/pyroelectric fields of the LN:Fe crystals. In this work, we report on the response of relatively large single droplets and groups of several microdroplets of a ferroelectric nematic LC material to photovoltaic fields on the surface of the X-cut and Z-cut LN:Fe plates. The field was generated by irradiation with a single green laser beam with a Gaussian intensity profile. The experiments were performed at different temperatures corresponding to the ferroelectric nematic ($N_F$), the standard nematic (N), and the isotropic (Iso) phases of the material. For comparison, we made similar experiments also with a conventional nematic LC material.

## 2. Experimental

The congruent LN:Fe crystals with 0.05 wt% dopant concentration were grown by using the Czochralski method. The details of the growth procedure are reported elsewhere [3]. The grown crystals were polarized, cut to 1.0 × 15.1 × 15.3 mm$^3$ (X × Y × Z; X-cut plate) and 15.3 × 15.1 × 1.0 mm$^3$ (X × Y × Z; Z-cut plate) pieces (Fig. 1(a)), and optically polished. Before the experiments, the plates were washed in acetone and dried by the flow of nitrogen.

The ferroelectric nematic liquid crystalline material used in the experiments (M5, Merck KGaA) exhibits the $N_F$ phase at room temperature. It undergoes the phase transition from the $N_F$ to an intermediate nematic $N_2$ phase (also known as splay phase) at 45°C, the transition from the $N_2$ to the N phase at 57°C, and the transition from the N to the Iso phase at 87°C. All LC phases of M5 are enantiotropic. For comparison, some experiments were also performed with the standard commercial nematic LC mixture (E7, Shijiazhuang Chengzhi Yonghua Display Material Co., Ltd.). At room temperature, E7 is in the N phase, and it experiences the phase transition to the Iso phase at 60°C.

For experiments at room temperature, a selected LN:Fe plate was placed onto a microscope glass slide and a few LC droplets with a volume of around 5 μl were deposited onto its top surface by the injection needle. Figures 1(b) and 1(c) show a side view of sessile droplets of E7 on the X-cut LN:Fe plate and of sessile droplets of M5 on the Z-cut LN:Fe plate, respectively. One can notice that the contact angles of both LC materials are practically the same, i.e., around 40°. Similar contact angles were also observed for E7 on the Z-cut plate and for M5 on the X-cut plate. For experiments at higher temperatures, the LN:Fe plate with the LC droplets was put into the microscope heating stage (STC200, Instec Inc.) and heated to the desired temperature. Before laser irradiation, the samples were left for 15 min at a fixed temperature to reach a stable state.

Optical irradiation of the assembly was implemented by a CW laser beam operating at the wavelength of 532 nm. The beam was directed onto the sample via the arm for the episcopic illumination in the polarization optical microscope (Optiphot-2-Pol, Nikon)(Fig. 1(d)). The spot size of the beam on the sample was 0.8 mm, and its optical power was 40 mW. The beam was impinging in the center of the viewing area. The irradiation-induced modifications of the LC droplet structure were monitored with polarization optical microscopy (POM) using the diascopic illumination configuration. The unwanted reflected laser light was filtered away by an appropriate notch filter. In most of the experiments, the angle between the polarizer and the analyzer was set to be 70°. This allowed us to resolve more details than in the case of crossed polarizers.

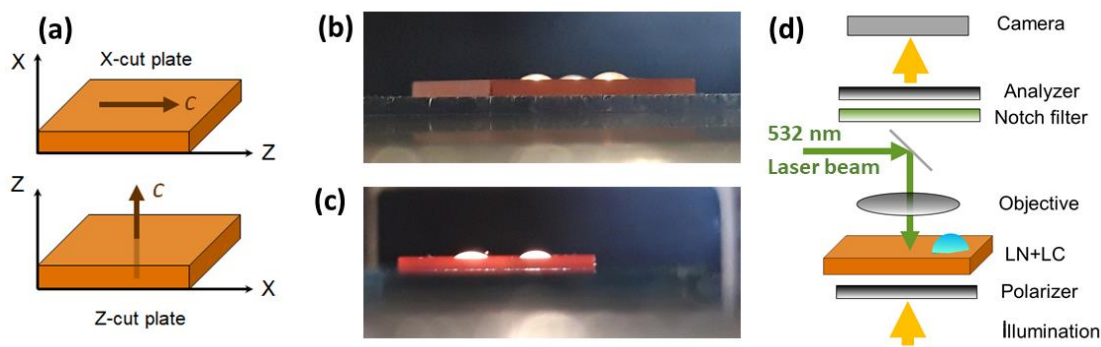

**Fig. 1.** (a) Schematic drawing of the X-cut and Z-cut LN:Fe plates used in the experiments. (b) Side-view image of sessile droplets of standard nematic LC material (E7) on X-cut plate at room temperature. (c) Side-view image of sessile droplets of ferroelectric nematic LC material (M5) on Z-cut plate at room temperature. (d) Schematic drawing of the setup for laser irradiation of the samples and monitoring of the induced modifications.

3. Results

*3.1. Experiments with single large droplets*

When the samples were illuminated so that the laser beam was impinging onto the LC droplet (see Fig. 1(c)), we observed only some relatively minor rearrangement of the internal orientational structure of the LC medium. Much more drastic effects occurred when the laser impacted outside the droplet area. Fig.2 shows four consecutive snapshots of the dynamic processes taking place at the edge of an M5 droplet deposited on the X-cut LN:Fe plate at room temperature, i.e, in the $N_F$ LC phase. The c-axis of the crystal is parallel to the vertical axis of the image. The green circle in Fig. 2(a) denotes the position of a laser beam with its center being located about 1 mm from the droplet edge. A few seconds after starting laser illumination, several tiny filaments are ejected from the droplet rim. They are oriented in the direction of the illuminated region (Fig. 2(b,c)). They are transient in nature and exhibit branching into narrower filaments. After dying out, they leave behind several tiny secondary droplets. As shown in Fig. 2(d), some filaments can reach lengths as large as several millimeters.

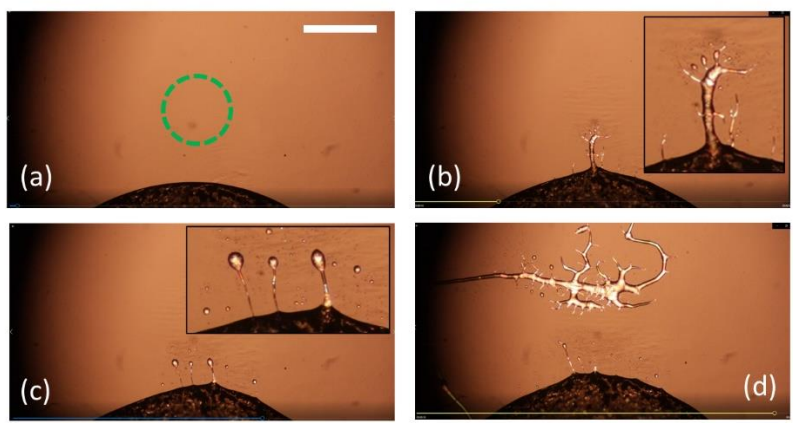

**Fig. 2.** Edge area of a large drop of M5 on top of X-cut LN:Fe at room temperature, i.e. in the $N_F$ phase. (a) At the starting of laser illumination, after (b) 18 s, (c) 47 s, and (d) 70 s. The location of a laser beam is indicated by the green circle in (a). The scale bar is 1 mm. The insets in (b) and (c) show enlarged parts of the images. (Full video is available in the supplementary material (video S1)).

The room temperature behavior of an M5 droplet placed on the the Z-cut LN:Fe plate is shown in Fig. 3. Also, in this case spikes and filaments start to protrude from the edge region of the droplet a few second after starting illumination. They are prone to branching and after retraction leave behind tiny secondary droplets.

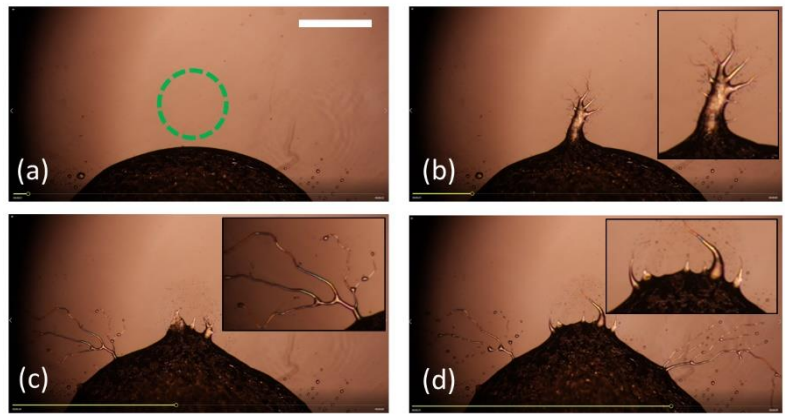

**Fig. 3.** Edge area of a large drop of M5 on top of Z-cut LN:Fe at room temperature, i.e. in the $N_F$ phase. (a) At the starting of laser illumination, after (b) 5 s, (c) 15 s, and (d) 25 s. The location of a laser beam is indicated by the green circle in (a). The scale bar is 1 mm. The insets in (b), (c) and (d) show enlarged parts of the images. (Full video is available in the supplementary material (video S2)).

The results of the same experiment with the droplet of the standard nematic LC material E7 placed on the X-cut LN:Fe plate are shown in Fig. 4. They reveal very different behavior from the one observed with the M5 droplets. Around 20 s after starting illumination, the E7 droplet edge begins to expand towards the illuminated area. The protrusion takes place in the form of a large rounded outgrowth that, after some time, partially retracts back and lefts behind some secondary droplets. The process then repeats again. The outgrowths do not show any branching and also no profound spikes.

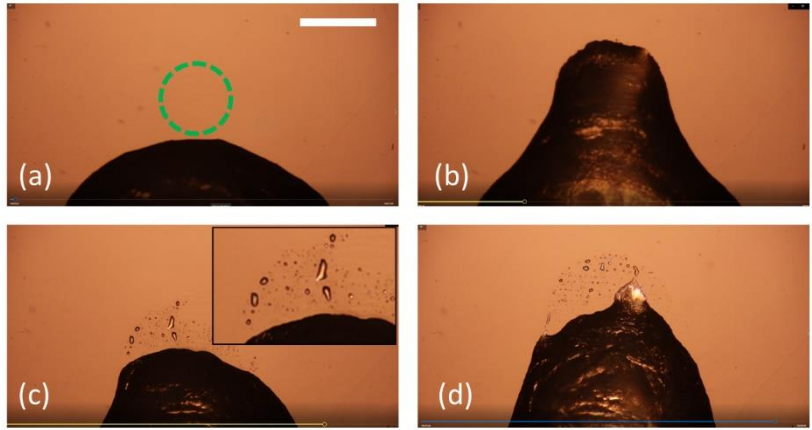

**Fig. 4.** Edge area of a large drop of E7 on top of X-cut LN:Fe at room temperature, i.e. in the N phase. (a) At the starting of laser illumination, after (b) 30 s, (c) 90 s, and (d) 105 s. The location of a laser beam is indicated by the green circle in (a). The scale bar is 1 mm. The inset in (c) shows an enlarged part of the image. (Full video is available in supplementary material (video S3)).

Similar behavior, but with less expansive protrusions, was observed also for the M5 droplet heated up to the nematic phase, i.e. to 75°C. The results obtained on the X-cut LN:Fe plate are shown in Fig. 5. The small droplets in the upper part of the image are remaining of the experiment performed at room temperature (see Fig.2(d)). Also, in this case, the temporary outgrowths do not show any branching and also no profound spikes.

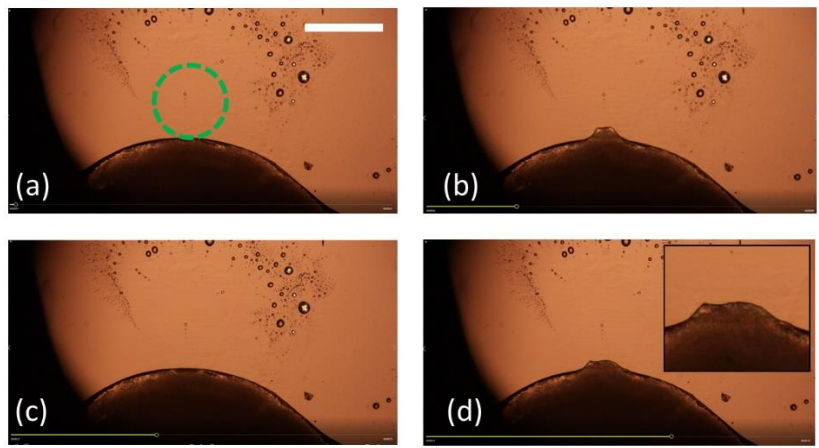

**Fig. 5.** Edge area of a large drop of M5 on top of X-cut LN:Fe at 75°C, i.e., in the N phase. (a) At the starting of laser illumination, after (b) 9 s, (c) 13 s, and (d) 27 s. The location of a laser beam is indicated by the green circle in (a). The scale bar is 1 mm. The inset in (d) shows an enlarged part of the image. (Full video is available in the supplementary material (video S4)).

### 3.2. Experiments with groups of small droplets

All the above-described experiments and also heating of the samples, in particular on the Z-cut LN:Fe plate, resulted in the creation of several small droplets. Therefore, we decided to investigate also what happens if laser illumination is performed in the area of several small droplets, i.e., droplets smaller than the spot size of the laser beam. Fig. 6 shows the results obtained with a group of M5 droplets on the X-cut LN:Fe at room temperature (in the $N_F$ phase).

The c-axis of the crystal is again parallel to the vertical axis of the image. Shortly after beginning of illumination, the droplets in the illuminated area get connected with tiny filaments. These connections are quite irregular and progress between the droplets in a spiky-like manner. With increasing time, they also expand to the droplets relatively far from the illuminated spot. The bridges persist as long as the illumination is present.

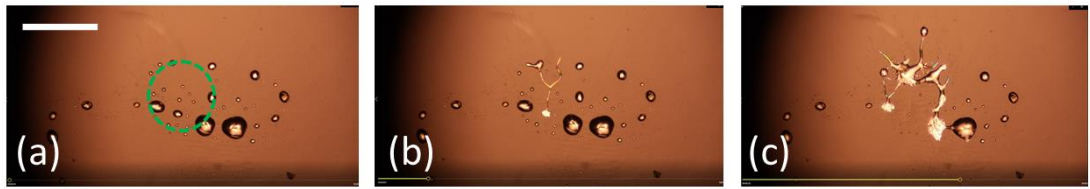

**Fig. 6.** A group of M5 droplets on X-cut LN:Fe at room temperature, i.e., in the $N_F$ phase. (a) At the starting of laser illumination, after (b) 5 s, and (c) 25 s. The location of a laser beam is indicated by the green circle in (a). The scale bar is 1 mm. (Full video is available in the supplementary material (video S5)).

The illumination-induced bridging between the droplets also takes place when the droplets are heated to 75°C, i.e., to the N phase. The results for a group of M5 droplets on X-cut LN-Fe are shown in Fig. 7. In contrast to the $N_F$ phase, the bridges formed in the N phase have a very short lifetime and they typically connect only near neighboring droplets. On the basis of this effect, some small droplets coalesce together into larger ones. A similar formation of very short-lived bridges was also observed when the droplets were heated to 100°C, i.e., in the isotropic phase (video is available in supplementary material as video S7).

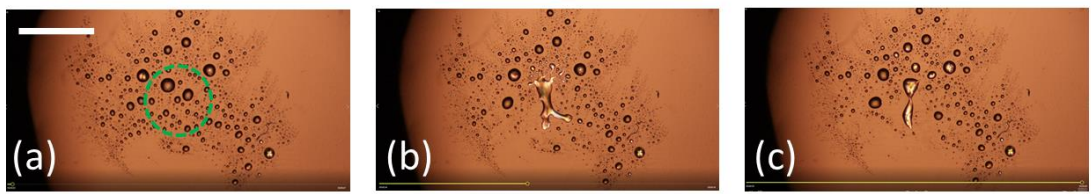

**Fig. 7.** The group of M5 droplets on X-cut LN:Fe at 75°C, i.e. in the N phase. (a) At the starting of laser illumination, after (b) 22 s, and (c) 50 s. The location of a laser beam is indicated by the green circle in (a). The scale bar is 1 mm. (Full video is available in supplementary material (video S6)).

The bridging effect can also be observed with the group of E7 droplets at room temperature. Two typical images are shown in Fig. 8. The filaments forming the bridges between the E7 droplets are more regular than for M5 droplets. On the X-cut LN:Fe plate they are localized to the illuminated area and preferentially oriented in the vertical direction – i.e. along the c-axis of the crystal. On the Z-cut plate they grow in the radial direction with respect to the illuminated area and expand quite far out from the illuminated region.

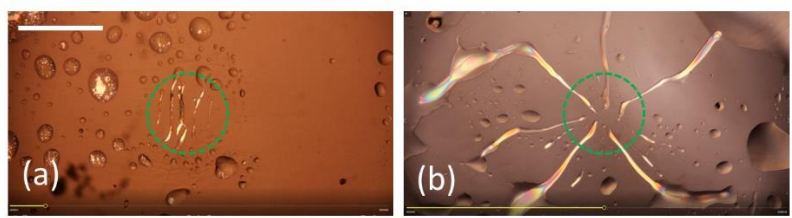

**Fig. 8.** The group of E7 droplets at room temperature during laser irradiation of the central region of the image. (a) On the X-cut LN:Fe there appear filaments bridging the droplets that are preferentially oriented in the vertical direction, i.e., along the c-axis of the crystal (b) On the Z-cut LN:Fe plate, the filaments grow out in the radial direction with respect to the illuminated area. The location of a laser beam is indicated by the green circles. The scale bar is 1 mm. (Full videos are available in the supplementary material (videos S8 and S9)).

Fig. 9. shows the behavior that can be attributed practically to a single small droplet of E7 located in the middle of the illuminated area on the X-cut LN:Fe plate. During illumination, the droplet transforms into tendrils that grow mainly along +c and -c directions (see Fig. 1(a)). After some time, the material of the initial droplet is transported into both edge regions of the illuminated area, where several smaller droplets are formed.

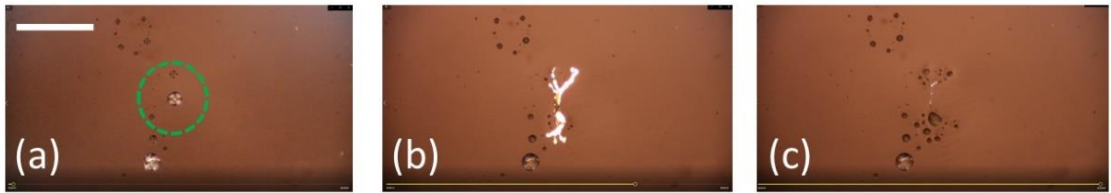

**Fig. 9.** A small E7 droplet on the X-cut LN-Fe plate positioned in the middle of the illuminated area during laser irradiation at room temperature. (a) At the starting of laser illumination, (b) 28 s afterward, (c) 41 s afterward. The location of a laser beam is indicated by the green circle in (a). The scale bar is 1 mm. (Full video is available in the supplementary material (video S10)).

## 4. Discussion

Laser illumination of LN:Fe crystals generates a photogalvanic current along the c-axis of the crystal. In the X-cut plate, this causes charge migration in the plane of the plate, and in the Z-cut plate, charges separate in the direction perpendicular to the plate [6,7]. In the open-circuit conditions, as used in our experiments, after some time, the current ends and stable spatial distribution of charges is generated. The associated static photovoltaic electric field **E**(**r**), which has a dipolar form, is localized to the illuminated area. The (di)electrophoretic force acting on surface objects located in and around the illuminated region is oriented predominantly along the c-axis. On the X-cut plate, this force is expected to push the objects along the c-axis, as observed in experiments with small LC droplets (Fig 7, Fig. 8(a), Fig. 9). On the Z-cut plate, the vertical component of the (di)electrophoretic force is compensated by gravity, so the dynamic processes are expected to be governed by the tangential component that exhibits a radial dependence with respect to the center of the illuminated area. This effect is clearly demonstrated in Fig. 8(b).

Our experiments reveal that, besides the $N_F$ phase, which is known to be extremely sensitive to the external electric field, also droplets of the conventional nematic (N) and even the isotropic phase show quite profound photovoltaic field-induced dynamic effects. The LC material forming single sessile droplets that are in size much larger than the illumination spot is dragged towards the illuminated region. The LC material forming droplets that are in size much smaller than the illuminated area, is rearranged into filamented structures in a symmetric manner with respect to the c-axis of the crystal. This observation suggests that dielectrophoretic forces dominate over the electrophoretic forces [40]. In the $N_F$ phase, the described processes take place via the rapid formation of conical spikes that often evolve into lengthy jet streams prone to multi-level branching (Fig 2(d), Fig 3(c)). The branching is, in our opinion, associated with the polar nature of the streams. In the N and the Iso phases, jetting effects are not present.

Analogous phenomena were widely investigated with droplets of charged and also uncharged conventional dielectric liquids, such as water, exposed to strong electric fields [41-44]. The spikes, also known as Taylor cones, and the associated jetting phenomena known as electrohydrodynamic tip-streaming, cone-jetting, Rayleigh jets, etc., are attributed to the competition between the (di)electrophoretic forces and surface tension of the liquid [40]. With increasing electric field, the uncharged droplets firstly stretch from spherical into spheroidal shape, and then Taylor cones (cusps) develop at their opposing ends. When the electric field riches some critical value $E_c$, deformed droplets become unstable, and cone-jetting begins. Similar phenomena also exist in (magnetic) ferrofluids exposed to strong magnetic fields, where they are associated with the Rosensweig instability [45-47].

The disintegration of cone-jets typically leads to the formation of tiny charged droplets, which is an important process in many technological applications, e.g., in jet-printing and electro-spraying [48]. Electrospray deposition is, among others, also used to fabricate nanodroplets of LC materials from their solution in suitable solvents [49]. Solvents are needed because the viscosity of LC materials is usually too high for direct electro-spraying processing. However, recently, Barboza et al. reported explosive jet-streaming phenomena observed in the $N_F$ phase of droplets of a ferroelectric nematic liquid crystalline compound (RM734 [34, 35]) exposed to the pyroelectric field of a Z-cut LN:Fe crystal [50]. This work revealed that the pyroelectric field created on the surface of LN:Fe crystals can be high enough to reach the critical field in the LC media.

Material parameters that determine the critical field associated with the transition of uncharged liquid droplets from stable Taylor cones to the cone-jetting effect are density, surface tension, viscosity, electric conductivity, and dielectric permittivity [40]. With changing the temperature of the LC medium to get different LC mesophases, the average values of these material parameters change with temperature, similarly to standard liquids. But, ferroelectric LC materials possess a very special property, namely that their (effective) dielectric permittivity in the $N_F$ phase is orders of magnitude larger than in the N phase. The values in the order of magnitude of $10^4$ are standardly reported [33, 51, 52]. We propose that this extraordinary increase of the dielectric permittivity is the main reason why the droplets of M5 exposed to the photovoltaic field in the $N_F$ phase exhibit intense spiking and jetting effects characteristic for $E>E_c$, while much less drastic features typical for $E<E_c$ are observed in the N phase of the same material and also in the N phase of a conventional nematic material. The origin of a "giant" dielectric permittivity of the $N_F$ phase is at present still debated.

Another specificity of the LC phases is that most of their material properties are strongly anisotropic, which is anticipated to generate diverse features with regard to the standard liquids, similar to the differences in photovoltaic trapping effects between isotropic and anisotropic nanoparticles [53]. The electrohydrodynamic behavior of anisotropic fluids in strong electric fields is, at present, still an open problem. We assume that the strong anisotropy of LC phases is the reason why even the standard nematic phase is much more responsive to the photovoltaic fields than the isotropic phase. Additional complexity in the depicted experiments is associated with the presence of the LN:Fe surface, on which the generated LC spikes and jets are advancing. Variations of surface adhesion and imperfections in surface topography can significantly affect the advancing processes. Also, the evanescent photovoltaic electric field is not homogeneous, but it varies in-plane in accordance with the illumination profile and rapidly decreases with the distance from the surface. Therefore, mainly only the bottom layer of the droplets in contact with the LN:Fe plate is exposed to the strong field-induced forces. This explains why smaller droplets are easier to be manipulated than bigger ones. All the above-mentioned features make theoretical modeling of the investigated phenomena very complicated.

## 5. Conclusion

Our preliminary observations of laser-illumination driven dynamics of LC droplets on the surface of LN:Fe crystals open up a complex electrohydrodynamic problem, to which they contribute much more questions than answers. Also, the presented results are far from being systematic and conclusive. An essential step forward will be to accomplish control over droplet generation and deposition processes so that one or more droplets of the same size can be placed onto (pre)selected positions on the LN:Fe plate. Another important improvement will be the usage of a fast camera, by which fine details of different stages of spike and jet generation can be resolved. To uncover the effect of different strengths of the photovoltaic field on the investigated phenomena, a set of LN:Fe plates with different concentrations of the dopant needs to be employed. In the future, such improvements can provide deeper insight into various electrohydrodynamic phenomena taking place when droplets of a LC material in different LC phases are exposed to strong external electric fields.

**Funding.** The work was supported by the Slovenian Research Agency (ARRS) in the framework of the research program P1-0192-Light&Matter, and by National Natural Science Foundation of China (12074201), PCSIRT (IRT 13R29). We are grateful to Merck for providing us with the ferronematic LC material (M5).

**Disclosures.** The authors declare no conflict of interest.

**Data availability.** The data used to support the findings of this study are available from the corresponding author upon request. The videomaterial is available in the supplementary material.

**Suplementary document.** See Suplement 1 for links to the videomaterial.